\documentstyle[preprint,eqsecnum,aps]{revtex}
\tighten
\begin{document}
\preprint{}
\draft
\title {The generalized Casimir operator\\
        and tensor representations of groups }
\author{V.D. Gladush\footnote{E-mail: gladush@ff.dsu.dp.ua},
            R.A. Konoplya }
\address{Department of Physics, Dnepropetrovsk State University, \\
per. Nauchny 13, Dnepropetrovsk, 320625 Ukraine}
\date{\today}
\maketitle
\begin{abstract}
There has been proposed a new method of the constructing of the basic
functions for spaces of tensor representations of the Lie groups with
the help of the generalized Casimir operator. In the definition of the
operator there were used
the Lie derivatives instead of the corresponding infenitisemal operators.
When introducing the generalized Casimir operator we use the metric for
which a group being considered will be isometry that follows
from the invariance condition for the generalized Casimir operator.
This allows us to formulate the eigenvalue and eigenfunction problems
correctly. The invariant  projection operators have been constructed in
order to separate irreducible components. The cases of the Bianchi type
$G^3IX$ and $G^3II$ groups are considered as examples.
\end{abstract}
\narrowtext
\section{INTRODUCTION}

   One of the ways to construct basic functions for the space of representation
of the group $G$ is to solve the eigenvalue problem for some complete set of
the invariant commutative Casimir operators.\cite{1,2}. These operators are
defined in the space of scalar functions. Note, that the Casimir operators
defined in a usual way in the space of functions with tensor values are not
invariant ones. To construct basic functions for the space of the tensor
representation of the group $G$, one decomposes objects, being considered,
into irreducible components, which, in their turn, are not the eigenfunctions
of the Casimir operator. Thus basic functions for the rotation group $SO(3)$
are the eigenfunctions of the Laplacian on two-sphere  $S^2$, although the
irreducible components of the tensors are eigenfunctions of the Laplacian on
the group $SU(2)$ \cite{3}.

 In the work \cite{Gl1} there was introduced  the generalized Casimir
operator $G$, which is invariant in the space of tensor functions. This made
it possible to formulate correctly and solve the eigenvalue and tensor
eigenfunction problems, and, thereby, to construct
tensor representations of various groups.
However, in this approach, the system of differential equations
appears with "tangled" components of tensor functions. That is
why the direct solving of the tensor eigenfunction problem,
generally speaking, is impossible. Moreover, due to the
decomposibility of the tensor representation into irreducible
ones, the spectrum of the generalized Casimir operator is, in the
general case, degenerate. Thus it is necessary to decompose
invariantly and to disentangle the mentioned system. It means that
we need to go over to the set of differential equations for
individual components or for their combinations. This reduces,
first, to the separating of the irreducible combinations of the
components for a given tensor field and to the classification of these
combinations \cite{In}, and, second, to the decomposition of the generalized
Casimir operator into irreducible parts.

    On the other hand in a number of works \cite{Gl2} - \cite{Gl-Kon}, the
authors constructed the theory of split-structures ${\cal H}^r$ on
the pseudo-Riemannian manifold $M^n$ and considered its
applications. The theory of split structures is a general approach
to the decomposition of the tangent bundle of pseudo-Riemannian manifolds
into $r$ subbundles
and the associated decomposition of geometric objects.
There have been considered split-structures
induced by groups of isometries and introduced the notion of
the split-structure compatible with a given group of isometry.
It turns out that the theory of split-structures on manifolds
give us an accurate and natural technique for solving of the
problems arising when finding the tensor eigenfunctions. The
technique makes it possible to "disentangle" the corresponding
system of differential equations, and, in fact, give us the
method for invariant decomposition of the generalized Casimir
operator into irreducible parts. In other words, the method
separates variables in a tensor differential equation  in
partial derivatives  generated by a given generalized Casimir
operator when dealing with the tensor eigenfunction problem.

    In the present work we consider applications of the notions
and methods developed in  \cite{Gl1} - \cite{Gl-Kon} to the construction of
tensor representations for groups acting on a given manifold
$M^n$ as groups of isometries.

    The method can be applied in physics to the invariant defining
and constructing of tensor multipoles (tensor harmonics) as well
as to the obtaining of expansions of tensor physical fields in terms
of multipoles, in fact, for any continuous groups. This finds its application
to the classification and computing of all linear perturbations of
a gravitational field for spaces of General Relativity with
determined symmetries.

    The work is organized in the following way. In Sec.II one
defines the generalized Casimir operator. The latter differs
from the usual Casimir operator. In the definition of the generalized
Casimir operator the Lie derivatives with respect to the infinitesimal
generators are used instead of the infinitesimal generators of a group. These
generators are the tangent vectors to the one-parameter subgroup,
or, in other words, are curves in $M^n$. The invariance condition
of the generalized Casimir operator yields the equation for  the metric
which is used in this operator's definition. Hence it follows that
the vectors being considered will be the Killing's vectors for
some metric. Then we formulate the eigenvalue and tensor
eigenfunction problems for the generalized Casimir operator.

    In Sec.III the problem of dioganalization of the generalized
Casimir operator is solved by using of a split-structure
${\cal H}^r$ compatible with the group $G^r$. The decomposition of the
corresponding system of differential equations for tensor
eigenfunctions into invariant irreducible differential subsystems
is constructed in order to disentangle the system. For the separating of
the irreducible components of tensor eigenfunctions we use the technique
of the invariant projection operators. As a result, the generalized
Casimir operator splits into the set of the invariant operators for
each irreducible component.

    In Sec.IV  the generalized Casimir operator are constructed for the
rotation group  $SO(3)$ in the three-dimensional Euclidean space.
The construction of the spherical tensor symmetric harmonics of type $(2,0)$
and weight $l$ is given as an instance.

    In Sec.V we deal with nonunitary representations for noncompact groups
$G^3~II$ according to the Bianchi classification. There is constructed
the generalized Casimir operator and given an example of the construction
of a point series of the representation.

\section{The generalized Casimir operator and
         tensor representations of groups}

    Let $M^n$ be an $n$-dimensional manifold and $G^r$ be an
$r$-parameter transformation group on $M^n$. For any differentiable vector
fields $X,Y,Z,\xi,e_a$ on $M^n$ the Lie bracket
$[X,Y]=-[Y,X]$ obeys the Jacobi identity
\begin{equation}\label{1}
   [X,[Y,Z]]+[Y,[Z,X]]+[Z,[X,Y]]=0.
\end{equation}
The Lie derivatives of a function $\varphi $, vector field $Y$, and
one-form  $\omega $ with respect to a vector $X$ are given by the
formulas
\begin{equation}\label{2}
{\cal L}_{X}\varphi = X\varphi ;~~~~L_{X}Y=[X,Y];~~~~
(L_{X}\omega)\cdot Z=X(\omega\cdot Z)-\\
   \omega\cdot [X,Z],
\end{equation}
where $\omega \cdot Z=\omega (Z)$ is an inner product of a
one-form $\omega $ and vector $Z$. The Lie derivative
$L_{X} \ T$ of a tensor $T$ of type (p,q) with respect to a vector
$X$ is given by
\begin{equation}\label{3}
(L_{X} \ T)(Y_1,...,Y_q) = L_{X} \ (T(Y_1,...,Y_q)) -
\sum^{q}_{i=1} T(Y_1,...,Y_{i-1},L_{X} \ Y_i,Y_{i+1},...,Y_q).
\end{equation}
The Lie algebra of the group $G^r$ is represented by the vector
fields $ {\xi_i}~(i=1,2,...,r)$, which are tangent to the
one-parameter subgroups and have the properties
\begin{equation}\label{3a}
   [ {\xi_i}, {\xi_j}] = C^k_{ij} {\xi_k} .
\end{equation}
Here $C^k_{ij}$ are the structure constants satisfying the condition
\begin{equation}\label{4}
   C^k_{ij}=-C^k_{ji};~~~~~C^p_{is}C^s_{jk}+C^p_{js}C^s_{ki}+\\
   C^p_{ks}C^s_{lj}=0 .
\end{equation}
Using the Jacobi identity we can easily find the relations for the commutators
of the Lie derivatives.
\begin{equation}\label{5}
[L_{ {\xi_i}},L_{ {\xi_k}}]=L_{[ {\xi_i}, {\xi_k}]}=C^j_{ik}L_{ {\xi_j}} ;
\end{equation}
\begin{equation}\label{6}
      [L_{ {\xi_i}},[L_{ {\xi_j}},L_{ {\xi_k}}]]+
      [L_{ {\xi_j}},[L_{ {\xi_k}},L_{ {\xi_i}}]]+
      [L_{ {\xi_k}},[L_{ {\xi_i}},L_{ {\xi_j}}]]=0.
\end{equation}
Thus the operators $L_{ {\xi_i}},~i=1,2,...r$ form
the representation of the Lie algebra for the group $G^r$. However, unlike
the Lie algebra of the tangent vectors $ {\xi_i}$, defined only on
functions, the Lie derivatives, defined on tensors of an arbitrary type
and rang, are invariant operators under the general coordinate
transformations. Hence we obtain the way of construction of the generalized
Casimir operator of the second order
\begin{equation}\label{7}
        G=g^{ik}L_{ {\xi_i}}L_{ {\xi_k}},
\end{equation}
where $g^{ik} \: (i,k=1,2,...r)$ are contravariant components of some unknown
ù metric, which is subject to be determined. By its definition this operator
commutes with all the operators $L_{ {\xi_i}}$ of the representation.
Hence it follows
\begin{equation}\label{8}
(L_{ {\xi_j}}g^{-1})^{ik}\equiv  {\xi_j} g^{ik}+C^i_{jl}g^{lk}+C^k_{jl}g^{il}=0 .
\end{equation}
Here the tensor
\begin{equation}\label{9}
g^{-1}=g^{ik} {\xi_i}\otimes {\xi_k}~~~~~(rang\,||g^{-1}||=s)
\end{equation}
defines the metric on covectors which belong to the surfaces of transitivity
 $M^s\subset M^n  (s\! \leq\! r,  ~s\! \leq\! n)$,where the symbol $\otimes $
denotes the tensor product. In accordance with (\ref{8}), the group $G^r$
on the surfaces of transitivity is a group of isometries, where $ {\xi_i}$
are the Killing's vectors. Solutions of the Killing's equations (\ref{8})
give us the metric for the generalized Casimir operator (\ref{7}).

It turns out that for semisimple groups that will be enough to consider
only constant solutions of (\ref{8}). Indeed, in this case the Cartan tensor
\begin{equation}\label{10}
        g_{ik}=\frac{1}{2}C^l_{ij}C^j_{lk},
\end{equation}
satisfying the equation
\begin{equation}\label{11}
        C^l_{ij}g_{lk}+C^l_{ik}g_{jl}=0
\end{equation}
is non-degenerate \cite{2}. Therefore there is the inverse of the tensor
$g^{ik}:(g^{ik}g_{kj}=\delta^{i}_{j})$, for which the conditions (\ref{11})
yield
\begin{equation}\label{12}
 C^i_{jl}g^{lk}+C^k_{jl}g^{il}=0.
\end{equation}
Comparing (\ref{12}) and (\ref{8}) we can conclude that $ {\xi}_jg^{ik}=0$
and $g^{ik}$ are constants on the surfaces of transitivity.

In the general case the consideration of the constant solutions of
(\ref{8}) will be inadequate for construction of the generalized Casimir
operator $G$, which is non-degenerate on $M^s$.

Let $T$ be a tensor of type $(p,q)$ on $M^s$. Then we have

\begin{equation}
T = T^{a_1...a_p}_{b_1...b_q}e_{a_1}\bigotimes...\bigotimes e_{a_p}
\bigotimes e^{b_1}\bigotimes ...\bigotimes e^{b_q},
\end{equation}
 where $e_a (a=1,2,...,s)$ is some vector basis on $M^s$, and
 $e^a,$~ $e^a(e_b) = \delta^a_b$,~ is a co-vector basis on $M^s$. Here
$T^{a_1...a_p}_{b_1...b_q}$ are the components of the tensor $T$ with respect
to the basis ${e_a}$. It will be the eigenfunction tensor of the generalized
Casimir operator $G$, provided the equation
\begin{equation}\label{13}
        GT \equiv g^{ik} L_{\xi_i} L_{\xi_k} T = \lambda T.
\end{equation}
is satisfied.
This equation can be rewritten in the following form
\begin{equation}\label{14}
{\cal G}T^{a_1...a_p}_{b_1...b_q} \equiv g^{ik}{\cal L}_{\xi_i}{\cal L}_{\xi_k}\\
T^{a_1...a_p}_{b_1...b_q}=\lambda T^{a_1...a_p}_{b_1...b_q},
\end{equation}
where ${\cal G}T^{a_1...a_p}_{b_1...b_q} \equiv (GT)^{a_1...a_p}_{b_1...b_q}$
are representations of  the generalized Casimir operator acting on the tensor
$T$ in the basis ${e_a}$. The representation of the Lie derivative
is determined by the formula
$${\cal L}_{\xi_i}T^{a_1...a_p}_{b_1...b_q}\equiv (L_{\xi_i}T)^{a_1...a_p}_{b_1...b_q}
= T^{a_1...a_p}_{b_1...b_q,c} \xi_i^c +
T^{a_1...a_p}_{c,b_2...b_q}\xi_{i,{b_1}}^c + \cdots$$
\begin{equation}
+ T^{a_1...a_p}_{b_1...b_{q-1}c}\xi_{i,{b_q}}^c -
T^{ca_2...a_p}_{b_1...b_q}\xi_{i,c}^{a_1} - \cdots -
T^{a_1...a_{p-1}c}_{b_1...b_q}\xi_{i,c}^{a_p}.
\end{equation}

It is difficult to solve the equation (\ref{14}) directly, because the
Lie operators "tangle" components of $T$.

\section{A split-structure on a manifold and diogonalization of the
         generalized Casimir operator}

In order to solve the system of equations (\ref{14}) it is necessary to
"disentangle" components of the tensor $T$ in  the equations, i.e. to
diogonalize the operator $G$. By means of decomposition the technique
of the dioganalization of the generalized Casimir operator $G$ can be
realized invariantly. Herewith the tensor equations (\ref{14}) split into the system
of scalar differential equations for irreducible components of the tensor
$T$. Now some additional definitions we used will be given below
\cite{Glpol}- \cite{Gl-Kon} .

  A linear operator $L$ on the tangent bundle $T(M)$ is a tensor of type
$(1,1)$ which acts according to the relation $L\cdot X\equiv L(X)\in T(M),~~\forall X\in T(M)$.
Then the formula
\begin{equation}
(L^{T}\cdot\omega)(X)=(\omega\cdot L)(X)\equiv\omega(L(X)),
~~~~~\forall X\in T(M)
\end{equation}
determines $L^{T}$, a  transpose of an operator $L$, which acts on
a one-form $\omega$.

The product of two linear operators $L\cdot H$ is defined by
$(L\cdot H)\cdot X=L\cdot(H\cdot X)\in T(M),~\forall X\in T(M)$.
An operator $H$ is called a symmetric one if
$$(H\cdot X,Y)=(X,H\cdot Y),~\forall X,Y\in T(M)$$.

We shall say that {\bf a split structure} ${\cal H}^s$ is introduced
on $M$ if the $s$ linear symmetric operators $H^a (a=1,2,...s)$ of a
constant rank with the properties
\begin{equation}\label{4a}
H^a\cdot H^b=\delta^{ab}H^b; ~~~~~~\sum_{a=1}^{s}H^a=I,
\end{equation}
where $I$ is the unit operator $(I\cdot X=I,~~\forall X\in T(M))$, are
defined on $T(M)$.

Then we can obtain the decomposition the tangent bundle $T(M)$ and
cotangent bundle $T^{*}(M)$ into the $(n_1+n_2+...+n_s)$ subbundles
$\Sigma^{a}$, $\Sigma_{a}^{*}$, so that
\begin{eqnarray} \label{7a}
    T(M) = {\bigoplus_{a=1}^{s}}\Sigma^{a};~~~~
T^{*}(M) = {\bigoplus_{a=1}^{s}}\Sigma^{*}_{a}\,.
\end{eqnarray}

Then arbitrary vectors, covectors, and metrics are decomposed according
to the scheme:
\begin{equation}\label{8a}
X = \sum^{s}_{a=1} X^a,~~~
\omega = \sum^{s}_{a=1}\omega_a,~~~
g = \sum^{s}_{a=1} g^a,~~~
g^{-1} = \sum^{s}_{a=1} g^{-1}_a
\end{equation}
where
\begin{equation}\label{9a}
X^a = H^a\cdot X,~~~
H^b\cdot X^a = 0,~~~
X^a\cdot X^b = 0,~~~(a\not= b)
\end{equation}
\begin{eqnarray}\label{10a}
     &\omega_a = \omega\cdot H^a,~~~
\omega_a (X^b) = 0,~~~(a\not= b).  \nonumber\\
\end{eqnarray}
   Using this scheme we can obtain the decomposition of more complex tensors.

   Let us now introduce an auxiliary definition. We shall say that a split
structure ${\cal H}^s$ is compatible with a group of isometries if the
conditions of invariance of ${\cal H}^s$ are satisfied, i.e. if
\begin{equation}\label{107}
{L}_{\xi_i }H^a = 0,~~~~ (i=1,2,...r; a=1,2,...s).
\end{equation}
The equations (\ref{4a}), (\ref{107}) define the invariant projection tensors.
The integrability conditions of (\ref{107}) are satisfied for solutions of
(\ref{107}) owing to (\ref{4a}).

In order to construct the projectors we require that there exist such
dual vector $\{e_a\}$ and covector $\{e^b\}$ bases on $M^s$, that
\begin{equation}\label{17}
e_a\cdot e^b =\delta^b_a~;
~~~H^a=e_a\otimes e^a .
\end{equation}
From now on we shall not sum on repeating indices $a$ and $b$. The invariance
condition of (\ref{107}) yields
\begin{equation}\label{18}
(L_{\xi_i}e^a)\cdot e_b=0
~~~~(a\! \neq \! b).
\end{equation}
Hence it follows
\begin{equation}\label{19}
        L_{\xi_i}e^a=
\mu^a_ie^a,
\end{equation}
where the factors of proportionality $\mu^{a}_{i}$ are some functions,
satisfying the equation
\begin{equation}\label{20}
        \xi_i\mu^a_k-\xi_k\mu^a_i=C^j_{ik}\mu^a_j ,
\end{equation}
which follows from the integrability condition of (\ref{19}). Using (\ref{2})
and (\ref{19}) we find
\begin{equation}\label{21}
        L_{\xi_i}e_a=-\mu^a_ie_a.
\end{equation}
Thus, the problem of the construction of the invariant projectors
reduces to the construction of the dual vector $\{e_a\}$ and
covector $\{e^b\}$ bases satisfying the system of equations
(\ref{19})-(\ref{21}). Some of the factors $\mu^a_i$, or even all
of them in some cases, can vanish. Then the projectors are
constructed by means of the invariant basis $\{e_a:
L_{\xi_i}e_a=0\}$. Thus in the case of a simply transitive group
$(r=s)$, the invariant vector basis $\{e_a\}$ can be expressed in
the form
\begin{equation}\label{22}
        e_a=L^b_a\xi_b~~~~~~(det||L^b_a||\neq 0).
\end{equation}
The factors $L^b_a$ satisfy the equations
\begin{equation}\label{23}
        \xi_bL^a_d+C^a_{bq}L^q_d=0 .
\end{equation}
The integrability conditions of these equations are satisfied owing to the
Jacobi identity.

If the invariant vector basis $\{e_a\}$ on $M^s$ is determined
then the inverse metric (~\ref{9}) in the case of the Riemannian
manifolds can be constructed by the formulas
\begin{equation}\label{24}
g^{-1}=\delta^{ab}e_a\otimes e_b=g^{ab}\xi_a\otimes \xi_b~;~~~~\\
g^{ab}=L^a_cL^b_d\delta^{cd}.
\end{equation}
Using (\ref{23}) it easily can be seen that the tensor $g^{-1}$, constructed
in accordance with (\ref{24}), actually satisfies the Killing's equations
(\ref{8}).

In any case of bases $\{e_a,e^b\}$ the initial
tensor $T$ can be expanded in the series
\begin{equation}\label{25}
  T=\sum^{ }_{{A,B}}\hat{T}^A_B=\sum^{ }_{{A,B}}T^A_B\hat{e}^B_A,
\end{equation}
where $\{\hat{e}^B_A\}=\{e_{a_1}\otimes\cdots\otimes e_{a_p}\otimes
e^{b_1}\otimes\cdots\otimes e^{b_q}\}$
is the tensor basis, $\hat{T}^{A}_{B} = T^{A}_{B}e^{B}_{A}$
is the tensor monomial and $T^A_B\equiv T^{a_1\ldots a_p}_{b_1\ldots b_q}$
is its component. $A=\{a_1,\ldots,a_p\}$ and $B=\{b_1,\ldots,b_q\}$ are
collective indices. The sum in (\ref{25}) comprises the complete set of indices
${A,B}$. It is easy to show that since the projectors $H_a$ are invariant,
the eigenvalue equations (\ref{13}) and (\ref{14}) split into the set of
independent eigenvalue invariant equations for monomials
\begin{eqnarray*}
G\hat{T}^A_B\equiv g^{ik}L_{\xi_i}L_{\xi_k}\hat{T}^A_B=\lambda\hat{T}^A_B.
\end{eqnarray*}
Using this relation together with (\ref{19}) and (\ref{20}) we obtain
\begin{equation}\label{26}
{\cal G}T^{A}_{B} \equiv g^{ik}{\cal L}_{\xi_i}{\cal L}_{\xi_k}T^{A}_{B}=\\
g^{ik}(\xi_i-\phi^{A}_{i B})(\xi_k-\phi^{A}_{k B})T^{A}_{B}=\\
\lambda T^A_B.
\end{equation}
Here
\begin{equation}\label{27}
\phi^A_{iB} = \sum^p_{k=1}\mu^{a_k}_i - \sum^q_{n=1}\mu^{b_n}_i,~~~
A=\{a_1,\ldots,a_p\},~~~ B=\{b_1,\ldots,b_q\}.
\end{equation}

Thus in order that the tensor equation (\ref{13}) could go over into the
invariantly split equations (\ref{26}) for the irreducible components
$T^A_B$, we must make a change
\begin{eqnarray*}
T\rightarrow T^A_B;~~~~~L_{\xi_i}\rightarrow{\cal L}_{\xi_i}=\xi_i-\phi^A_{i B}.
\end{eqnarray*}
The equations (\ref{26}) can be rewritten in the form
\begin{equation}\label{28}
{\cal G}T^{A}_{B} = [K-2g^{ik}\phi^A_{iB}\xi_k-g^{ik}\xi_i\phi^ A_{kB} +\\
g^{ik}\phi^A_{iB}\phi^A_{kB}]T^A_B=\lambda T^A_B,
\end{equation}
where
\begin{equation}\label{29}
K = g^{ik}\xi_i\xi_k
\end{equation}
is the standard Casimir operator defined in the space of scalar functions.
The solutions of the equations (\ref{28}) and (\ref{19}), (\ref{20}) give us
the basic tensor functions $\hat{T^A_B}=T^A_B\hat{e^B_A}$ in the space of a
tensor representation of the group $G^r$ (or, in other words, tensor
harmonics). Note that if there is the invariant basis (\ref{22}), then the
generalized Casimir operator (\ref{13}) with respect to this basis reduces
to the standard Casimir operator $K$, and in order to construct the tensor
basis of representation that will be enough to determine the basis of representation
in the space of scalar functions.

\section{The generalized Casimir operator for the rotation group
         and its tensor representations}

  Let us consider, as an example and comparison with the known results,
the case of the rotation group $SO(3)$ in the three-dimensional Euclidean
space. The Lie algebra of this group in terms of the spherical coordinate system
is represented by the following tangent vectors \cite{4}:
\begin{equation}\label{30}
\xi_1=\sin\varphi\frac{\partial
}{\partial\theta}+\cot\theta\cos\varphi\\ \frac{\partial
}{\partial\varphi}\, ; \quad \xi_2=-\cos\varphi\frac{\partial
}{\partial\theta}+\cot\theta\sin\varphi\\ \frac{\partial
}{\partial\varphi}\, ; \quad \xi_3=-\frac{\partial
}{\partial\varphi}.
\end{equation}
The Lie bracket is $[\xi_i,\xi_j]=\varepsilon_{ijk}\xi_k$ ,
where $\varepsilon_{ijk}$ are the Levi-Civita's symbols. As a usual one
should go over into the vectors generating the creation and annihilation
operators.
\begin{equation}\label{31}
H_s=-s\xi_2 + \imath \xi_1
 =e^{\imath s\varphi}\left(s\frac{\partial
}{\partial\theta}+\imath\cot\,\theta\\ \frac{\partial
}{\partial\varphi}\right)\, ; \qquad H_3=-\imath\frac{\partial
}{\partial\varphi}\, ; \qquad s=\pm 1.
\end{equation}

In the case being considered the surfaces of transitivity $M^2=S^2:r=const$
are two-dimensional, although the space of the group is three-dimensional.
The Cartan tensor (\ref{10}) is non-degenerate in this case, and
$g_{ik}=g^{ik}=\delta_{ik}$. Therefore we can take the operator
\begin{equation}\label{32}
K=\xi^2_1+\xi^2_2+\xi^2_3=-(H_{+1}H_{-1}+H^2_3-H_3)=\\
\frac{1}{\sin\theta}\frac{\partial }{\partial\theta}\\
\left(\sin\theta\frac{\partial }{\partial\theta}\right)+\frac{1}{\sin^2\theta}\\
\frac{\partial^2}{\partial\varphi^2}.
\end{equation}
as the standard Casimir operator (\ref{29}). Commutation relations for the Lie
operator generated by the vectors (\ref{31}) have the form
\begin{equation}\label{33}
[ L_{H_S},L_{H_3} ] = -sL_{H_S}\, ,~~(s=\pm 1); \qquad [ L_{H_{+1}},L_{H_{-1}} ] = 2L_{H_3}.
\end{equation}
The operators $L_{H_S}$ are the creation and annihilation operators for tensor
functions $\hat{T^A_B}$, which, in their turn, are the eigenfunctions of
the operator $L_{H_3}$. Following  \cite{4} we can show that there is the
set of tensor functions $T^{(l) A}_{(m) B}$ for which
\begin{equation}\label{34}
{\cal L}_{H_3}T^{(l) A}_{(m) B} = mT^{(l) A}_{(m) B} \qquad
      (m=-l,-l+1,\ldots,l)\, ,
\end{equation}
\begin{equation}\label{35}
{\cal L}_{H_S}T^{(l) A}_{(m) B} = \sqrt{l(l+1)-m(m+s)}T^{(l) A}_{(m) B}~,
\end{equation}
where $l$ is the weight of the representation. Herewith the tensor eigenfunction
equations (\ref{14}) can be written in the form
\begin{equation}\label{36}
-{\cal G}T^{(l) A}_{(m) B}=
({\cal L}_{H_{+1}}{\cal L}_{H_{-1}}+{\cal L}_{H_3}{\cal L}_{H_3}-{\cal L}_{H_3})\\
T^{(l) A}_{(m) B}=l(l+1)T^{(l) A}_{(m) B}.
\end{equation}

  Suppose that we need to consider the spherical tensor symmetric harmonics of
type $(2,0)$ and of weight $l$, which we shall denote $T^l$. With respect to
the initial differential basis
$\{e^r,~~dx^1=d\theta,~~dx^2=d\varphi\}$
they can be written in the form
\begin{equation}\label{37}
T^l=T^{(l)}_{rr}e^r\otimes e^r
+T^{(l)}_{ra}(e^r\otimes ~dx^a+dx^a\otimes e^r)\\
+T^{(l)}_{ab}dx^a\otimes dx^b.
\end{equation}
Note, that the covector $e^r=dr$
is invariant with respect to the rotation group. In order to split the system of
equations (\ref{36}) for the tensor (\ref{37}), into the irreducible
components, one should go over into the basis of one-forms on the surfaces of
transitivity $S^2$ satisfying the condition (\ref{19}).
It turns out that the covectors
\begin{equation}\label{38}
e^s =d\theta+\imath s\sin\theta d\varphi~~~~(s=\pm 1)
\end{equation}
are required. Herewith, the condition of invariance of a split structure
(\ref{107}) for the Lie operators associated with the vectors (\ref{31})
stipulates the relation
\begin{equation}\label{39}
\mu^s_{s'}=\frac{se^{\imath s'\varphi}}{\sin\theta}\, , \quad \mu^s_3=0
      \qquad (s,s'=\pm 1).
\end{equation}
By using (\ref{38}) the relation (\ref{37}) can be rewritten in the form
\begin{equation}\label{40}
T^l=T^{(l)}_{rr}e^r\otimes e^r+T^{(l)}_{rs}(e^r \otimes e^s
+e^s \otimes e^r)\\
+T^{(l)}_{ss'}e^s \otimes e^{s'}.
\end{equation}
where the sum on $s=\pm 1$ is implied. Then, if we suppose
\begin{equation}\label{41}
T^{(l)}_{rr}=h_{rr}t^l\, ; \quad T^{(l)}_{rs}=h_rt^l_s\, ; \quad
T^{(l)}_{rs}=ht^l_{s+s'}\, ,
\end{equation}
where $h_{rr},~h_r,~h$ are functions of $r$, then for the function
$t^l_n~~(n=0,..., s,..., s+s')$ we obtain
\begin{equation}\label{42}
\left\{\frac{1}{\sin\theta}\frac{\partial
}{\partial\theta}\sin\theta\\ \frac{\partial
}{\partial\theta}+\frac{1}{\sin^2\theta}\\
\left(\frac{\partial^2}{\partial\varphi^2}-2\imath\,
n\cos\theta\frac{\partial }{\partial\varphi}-n^2\right)\\
+l(l+1)\right\}t^l_n=0.
\end{equation}
Owing to (\ref{34}) we find
\begin{equation}\label{43}
t^l_n=e^{\imath m\varphi}P^l_{nm},
\end{equation}
where the functions $P^l_{nm}$ satisfy the ordinary differential equation
following from (\ref{42}):
\begin{equation}\label{44}
\left\{\frac{1}{\sin\theta}\frac{d}{d\theta}\sin\theta\\
\frac{d}{d\theta}-\frac{m^2-2mn\cos\theta+n^2}{\sin^2\theta}\\
+l(l+1)\right\}P^l_{nm}=0.
\end{equation}
The solutions of the obtained equations are the functions
$P^l_{nm}(\cos\theta)$ \cite{Land}, which appear in the theory of representation
of the rotation group  and are proportional to the Jacobi polynomials
$P^{(\alpha,\beta)}_k$. Recurrent relations for $P^l_{nm}$ follow from
(\ref{34}),(\ref{35}).

\section{The generalized Casimir operator for the Bianchi type
         $G^3~II$ group and its tensor representations}

  Let us consider, briefly, the case of nonunitary representations for
noncompact groups. We shall take as an example the three-parameter
non-Abelian group acting on $M^3$ with the coordinates $\{x,y,z\}$, i.e.
$G^3 ~II$  according to the Bianchi classification \cite{11}. The Lie
algebra of this group is represented by the vectors
\begin{equation}\label{45}
\xi_1=\frac{\partial }{\partial x}\, ; \qquad
\xi_2=x\frac{\partial }{\partial x}+
\frac{\partial }{\partial y}\, ; \qquad
\xi_3=\frac{\partial }{\partial z}\, ; \qquad
[\xi_1,\xi_2]=\xi_1\, .
\end{equation}
If we, for simplicity, make a substitution
$(x,y,z)\rightarrow (\upsilon =xe^{-y},y,z)$,
then
\begin{equation}\label{46}
\xi_1=e^{-y}\frac{\partial}{\partial\upsilon}\, ; \qquad
\xi_2=\frac{\partial}{\partial y}\, ; \qquad
\xi_3=\frac{\partial }{\partial z}\, ; \qquad [\xi_1,\xi_2] = \xi_1\, .
\end{equation}
Since the Cartan tensor (\ref{10}) is degenerate in this case, then in
order to find the non-degenerate metric on $M^3$ it is necessary that the
Killing's equations (\ref{8}) be solved at $C^1_{12}=1~;~C^2_{12}=0~;~
C^3_{12}=0$. This can easily be done by the constructing of the invariant
vector basis $\{e_a\}$, which satisfies the equations
$L_{\xi_a}e_b=0$ . The solution of the equation of the
invariance can be expressed in the form
\begin{equation}\label{47}
e_1 = \frac{\partial}{\partial\upsilon} = e^y \xi_1\, ; \qquad
e_2=\partial y-\upsilon\frac{\partial}{\partial\upsilon}
= \xi_2-\upsilon e^y \xi_1\, ; \qquad e_3 = \xi_3\, .
\end{equation}
The required metric, according to (\ref{24}), is written in the form
\begin{eqnarray}\label{}
&  g^{-1}=e_1 \otimes e_1+e_2 \otimes e_2+e_3 \otimes e_3 \nonumber \\
&  = (1+\upsilon^2)e^{2y}\xi_1 \otimes \xi_1
-\upsilon e^y (\xi_1 \otimes \xi_2+\xi_2 \otimes \xi_1)+
\xi_2 \otimes \xi_2+\xi_3 \otimes \xi_3\, .  \label{48}
\end{eqnarray}
Hence, omitting the symbol of the tensor product, we obtain the generalized
Casimir operator.

In the case being considered the finding of the tensor harmonics
$\hat{T}^{(\lambda)A}_{(\mu\nu)B}=T^{(\lambda)A}_{(\mu\nu)B}\hat{e}^B_A,~$ where
$T^{(\lambda)A}_{(\mu\nu)B}=h^A_Bt^{\lambda}_{\mu\nu}$, reduces to the
construction of the scalar harmonics $t^{\lambda}_{\mu\nu}$. The latter
are the eigenfunctions of the operators
\begin{equation}\label{49}
\xi_2 t^{\lambda}_{\mu\nu}=\frac{\partial }{\partial y} t^{\lambda}_{\mu\nu}\\
=\mu t^{\lambda}_{\mu\nu},
\end{equation}
\begin{equation}\label{49 1}
\xi_3 t^{\lambda}_{\mu\nu}=\frac{\partial }{\partial z} t^{\lambda}_{\mu\nu}\\
=\nu t^{\lambda}_{\mu\nu},
\end{equation}
\begin{equation}\label{50}
Kt^{\lambda}_{\mu\nu}~\equiv~\left[(1+\upsilon^2)\frac{\partial^2}{\partial\upsilon^2}\\
-2\upsilon\frac{\partial^2}{\partial\upsilon\partial y}+\\
\upsilon\frac{\partial}{\partial\upsilon}\\
+\frac{\partial^2}{\partial y^2}+\frac{\partial^2}{\partial z^2}\right]t^{\lambda}_{\mu\nu}=\lambda t^{\lambda}_{\mu\nu}.
\end{equation}
where $\mu$ and $\nu$ are  the eigenvalues of the operators $\xi_2$ and
$\xi_3$ respectively. These relations are analogies of the equations (\ref{34}), (\ref{36}) for the
group $SO(3)$. From the equations (\ref{49}), (\ref{49 1}) one finds
\begin{equation}\label{51}
t^{\lambda}_{\mu\nu}=f^{\lambda}_{\mu\nu}e^{\mu y+\nu z},
\end{equation}
where $f^{\lambda}_{\mu\nu}=f^{\lambda}_{\mu\nu}(\upsilon)$ satisfies the ordinary
differential equation that follows from (\ref{50})
\begin{equation}\label{52}
\left[(1+\upsilon^2)\frac{d^2}{d\upsilon^2}+(1-2\mu)\upsilon\frac{d}{d\upsilon}\\
+\mu^2+\nu^2\right]f^{\lambda}_{\mu\nu}=\lambda f^{\lambda}_{\mu\nu}.
\end{equation}
Since the index $\nu$ comes algebraically into (\ref{52}) we can rewrite the
equation (\ref{52}) in the following form:
\begin{equation}\label{201}
\left[(1+\upsilon^2)\frac{d^2}{d\upsilon^2}+(1-2\mu)\upsilon\frac{d}{d\upsilon}\\
+\mu^2\right]\varphi^{\sigma}_{\mu}=\sigma\varphi^{\sigma}_{\mu},
\end{equation}
where we used the substitution:
$f^{\lambda}_{\mu\nu}=A^{\lambda}_{\mu\nu}\varphi^{\sigma}_{\mu},~~
~~ A^{\lambda}_{\mu\nu}=const,~~ \sigma=\lambda-\nu^2$. The
general solution of (\ref{201}) can be written in  the form
$$\varphi(v)=A\, _{_2}F_{1}([-\mu-\sqrt{2\mu^2-\sigma}]/2,
[-\mu+\sqrt{2\mu^2-\sigma}]/2,1/2,-v^2)$$
\begin{equation}\label{202}
+\sqrt{v^2}B\, _{_2}F_{1}([-1-\mu-\sqrt{2\mu^2-\sigma}]/2,
[1-\mu+\sqrt{2\mu^2-\sigma}]/2,3/2,-v^2),
\end{equation}
where $A$ and $B$ are constant, and the hypergeometric function $_{_2}F_{1}(a,b,c,z)$ is the solution
of the following equation:
\begin{equation}\label{203}
z(1-z)\frac{d^{2}F}{dz^2}+[c-(a+b+1)z]\frac{dF}{dz}-abF=0.
\end{equation}

It is easy to show from the  commutation relations (\ref{46}) that
\begin{equation}\label{53}
t^{\lambda}_{\mu -1,\nu}=\xi_1 t^{\lambda}_{\mu\nu}.
\end{equation}
Hence it follows
\begin{equation}\label{54}
\frac{d}{d\upsilon}f^{\lambda}_{\mu\nu}=f^{\lambda}_{\mu -1,\nu}.
\end{equation}
Since the group $G^3~II$ is noncompact, then the spectra of the operators
(\ref{49}),(\ref{50}) are continuous. Here we shall consider, as one more
example, point series of the representation of $G^3~II$. For this purpose
we shall take $\sigma=n^2$. Then for the case of $\mu =n$ from (\ref{201})
we obtain
\begin{equation}\label{55}
\varphi^{n^2}_n~=~C\int (1+\upsilon^2)^{n-1/2}\,d\upsilon,
\end{equation}
where $C$ is the constant of integration (an additive constant is omitted).
Applying the relation (\ref{54}) to (\ref{55}) $n-m-1$ times one has
\begin{equation}\label{56}
\varphi^{n^2}_m~=~C\frac{d^{n-m-1}}{d\upsilon^{n-m-1}}(1+\upsilon^2)^{n-1/2}~~~\\
(m<n).
\end{equation}
Then we immediately obtain the particular solutions of the equations
(\ref{52}) $f^{\lambda}_{\mu\nu}$ as
$A^{\lambda}_{\mu\nu}\varphi^{\sigma}_{\mu}$ when $\sigma = n^2,~ \mu
=n,~\lambda = \nu^{2}+n^2$. That this relation is really the solution of the
equation (\ref{52}) one can easily verify by the
differentiating of the equation (\ref{52}) $n-m-1$ times under initial
values $\sigma=\mu^2 =n^2$.

   In the conclusion we shall write, for instance, the tensor eigenfunctions
of the generalized Casimir operator $G$ and of the Lie operator
${\cal L}_{\xi_2}$ for the covector harmonic $A^{(\lambda)}_{(\mu\nu)}$.
Using the invariant basis of the one-forms
\begin{eqnarray*}
e^1 = d\upsilon +\upsilon\,
dy = e^{-y}dx~;~~e^2=dy~,
~e^3=dz~
\end{eqnarray*}
which is dual to the basis (\ref{47}), we find
\begin{equation}\label{57}
A^{(\lambda)}_{(\mu\nu)}=\left[a^{(\lambda)}_{(\mu\nu)1}e^{-y}dx+
a^{(\lambda)}_{(\mu\nu)2}dy+a^{(\lambda)}_{(\mu\nu)3}dz\right]
t^\lambda_{\mu\nu}
\end{equation}
where
\begin{equation}\label{58}
t^{n^2}_{m\nu}~=~e^{ym+z\nu}\frac{d^{n-m-1}}{d\upsilon^{n-m-1}}(1+\upsilon{^2})^{n-1/2};~~~\\
\upsilon = xe^{-y},
\end{equation}
$a^{(\lambda)}_{(\mu)(\nu)1},~a^{(\lambda)}_{(\mu)(\nu)2},
~a^{(\lambda)}_{(\mu)(\nu)3}$  are constant.

\section{Conclusion}
In such a way, the possibilities of our method have been considered for
the Bianchi type $G^3~IX=SO(3)$ and $G^3~II$ groups.
Nonetheless it is evident that  the present method, in fact, makes
it possible to construct tensor representations of any continuous group $G$.

The question remains is whether it is possible to generalize this
method not only for tensor but also for spinor fields. The notion
of the Lie derivative of spinor fields was introduced by I.~Kosmann
(see \cite{Kosm}). In  \cite{Bilya} by extending the spinor representation
of the Lorentz group to the representation of the general linear group
$GL(4)$ the spinor fields are considered in arbitrary frames, and
thus there were defined the Lie derivatives of the spinor fields
with respect to an arbitrary vector field. Recently a geometric
definition of the Lie derivative for spinor fields, more general
than Kosmann's one, has been proposed in \cite{Godina}. When choosing
special infinitesimal lift (namely, for Kosmann vector fields) their
definition coincides with that given by I.Kosmann.

The essential property we used in our method is that  the
commutator of the Lie derivatives of tensor fields with respect to
vector fields equals to the Lie derivatives of tensor fields with
respect to the commutator of the vector fields. However for spinor fields,
as it follows from \cite{Kosm},\cite{Bilya}, this is not so. In this
connection it seems alluring to give a  new definition (if it is possible)
of the Lie derivative for spinor fields that will satisfy this requirement.


\def\CMPh{Commun. Math. Phys.}
\def\JPh{J. Phys.}
\def\CJP{Czech. J. Phys.}
\def\LMPh {Lett. Math. Phys.}
\def\NPh  {Nucl. Phys.}
\def\PhE  {Phys.Essays}
\def\PhL  {{\it Phys. Lett.}~}
\def\PhR  {{\it Phys. Rev.}~}
\def\PhRL {Phys. Rev. Lett.}
\def\PhRp {Phys. Rep.}
\def\NCim {Nuovo Cimento}
\def\NuPB {Nucl. Phys.}
\def\GRG {{\it Gen. Relativ. Gravit.}~}
\def\CQG {Class. Quantum Grav.}
\def\prp {report}
\def\Prp {Report}
\def\GrC {{\it Gravitation$\&$Cosmology}~}
\def\DANS {{\it Dokl.Akad.Nauk SSSR}~}
\def\APh {{\it Ann.Phys.}~}
\def\JMM {{\it Journ.Math. and Mech.}~}
\def\JMP {{\it J.Math. Phys.}~}
\def\IVUZ {{\it Izv.Vyssh.Uchebn.Zaved.Fiz}~}
\def\APP {{\it Acta Phys.Pol.}~}
\def\UFZh {{\it  Ukr. Fiz. Zh.}~}
\def\TMF {{\it Teor. i Mat. Fiz.}~}

\def\jn#1#2#3#4#5{{#1}{#2} {\bf #3}, {#4} {(#5)}}

\def\boo#1#2#3#4#5{{\it #1} ({#2}, {#3}, {#4}){#5}}

\def\prpr#1#2#3#4#5{{``#1,''} {#2 }{#3}{#4}, {#5} (unpublished)}


\begin{thebibliography}{99}
\bibitem{In}{In fact, it will be the construction of the irreducible
            representation of the group , but, that is just what we avoid.}
\end{thebibliography}
\end{document}